\documentstyle[11pt,epsfig]{article}
\input{epsf}
\begin{document}
 
\begin{center}
{\Large \bf Effects of Environment and Energy Injection on Gamma-Ray Burst 
Afterglows\footnote{To appear in the proceedings of the First KIAS International 
Workshop on Astrophysics: Explosive Phenomena in Astrophysical Compact Objects, 
Seoul, Korea; 24-27 May 2000.}}
 \vskip 0.5cm
{{\large Z. G. Dai} }
\vskip 0.2cm
{\em Department of Astronomy, Nanjing University, Nanjing 210093, China}
\end{center}

\centerline{\bf Abstract} 
\vskip 0.5cm 
There is growing evidence that some long gamma-ray bursts (GRBs) arise from the 
core collapse of massive stars, and thus it is inevitable that the environments of 
these GRBs are preburst stellar winds or dense media. We studied, for the first 
time, the wind model for afterglows based on the Blandford-McKee  self-similar 
solution of a relativistic shock, and suggested that GRB 970616 is an interactor 
with a stellar wind. We also proposed a dense medium model for some afterglows, 
e.g., the steepening in the light curve of the R-band afterglow of GRB 990123 
may be caused by the adiabatic shock which has evolved from an ultrarelativistic 
phase to a nonrelativistic phase in a dense medium. We further discussed the dense 
medium model in more details, and investigated the effects of 
synchrotron self absorption and energy injection. A shock in a dense medium
becomes nonrelativistic rapidly after a short relativistic phase.
The afterglow from the shock at the nonrelativistic stage decays more
rapidly than at the relativistic stage. Since some models for GRB energy sources 
predicted that a strongly magnetic millisecond pulsar may be born during GRB 
formation, we discussed the effect of such a pulsar on the evolution of the 
nonrelativistic shock through magnetic dipole radiation. We found that in the 
pulsar energy injection case, the dense medium model fits very well all the 
observational data of GRB 980519.  Recently, we combined the dense 
medium model with the pulsar energy injection effect to provide a good fit 
to the optical afterglow data of GRB 000301C.


\section{Introduction}
 
In the standard afterglow shock model (for a review see [27,38]),
a gamma-ray burst (GRB) afterglow is usually believed to be produced
by synchrotron radiation or inverse Compton scattering in
an ultrarelativistic shock wave expanding in a homogeneous medium.
As more and more ambient matter is swept up, the shock gradually
decelerates while the emission from such a shock fades down, dominating
at the beginning in X-rays and progressively at optical to radio energy
band.  The standard model is based on four basic assumptions:
(1) the total energy of the shock is released impulsively before its
formation; (2) the medium swept up by the shock is homogeneous
and its density ($n$) is the one of the interstellar medium
$\sim 1\,{\rm cm}^{-3}$; (3) the electron and magnetic field energy
fractions of the shocked medium and the index ($p$) in the accelerated
electrons' power-law distribution are constant during the whole evolution
stage; and (4) the shock is spherical.  

Each of these assumptions has been varied to discuss why some observed
afterglows deviate from that expected by the standard afterglow model.
For example, the R-band light curve of GRB 970508 afterglow peaks
around two days after the burst, and there is a rather rapid rise before
the peak which is followed by a long power-law decay. There are two models
explaining this special feature: (i) It was envisioned [28] 
that a postburst fireball may contain shells with a continuous
distribution of Lorentz factors. As the external forward shock sweeps up
ambient matter and decelerates, internal shells will catch up with
the shock and supply energy into it. A detailed calculation shows that
this model can explain well this special feature [26]. (ii) We considered 
continuous energy injection from a strongly magnetized millisecond pulsar into
the shock through magnetic dipole radiation [8]. This model can also
account for well the observations. It is very clear that these models
don't use basic assumption (1).

There are several models in the literature that discuss the effect of
inhomogeneous media on afterglows [9,23,3,4], dropping the second assumption.
Generally, an $n\propto r^{-k}$ ($k>0$) medium is expected to steepen
an afterglow's temporal decay. We studied, for the first time, 
the wind model for afterglows based on the Blandford-McKee  self-similar 
solution [1] of a relativistic adiabatic shock, and suggested that GRB 970616
is an interactor with a stellar wind of $n\propto r^{-2}$ [9]. It was found [3]
that a Wolf-Rayet star wind likely leads to an $n\propto r^{-2}$ medium,
and thus if GRB 980519 resulted from the explosion of such a massive star,
subsequent evolution of a relativistic shock in this medium is 
consistent with the steep decay in the R-band light curve of the afterglow
from this burst. Another way of dropping the second assumption is
that the density of an ambient medium is invoked to be as high as
$n\sim 10^6\,{\rm cm}^{-3}$. The temporal decay of the R-band afterglow 
of GRB 990123 has been detected to steepen about 2.5 days
after this burst [21,2,16]. We proposed
a plausible model in which a shock expanding in a dense medium
has evolved from a relativistic phase to a nonrelativistic phase [11].
We found that this model fits well the observational data if the medium
density is about $3\times 10^6\,{\rm cm}^{-3}$. We further suggested
that such a medium could be a supernova or supranova or hypernova ejecta.

In basic assumption (3), the electron and magnetic field energy fractions
of the shocked medium may not be varied during whole evolution,
as argued in [34], where all the observational data including both the prompt 
optical flash and the afterglow of GRB 990123 were analyzed.  

The steepening in the light curves of the afterglows of some 
bursts may also be due to lateral spreading of a jet, as analyzed
in [29,31] when the jet expands in a homogeneous interstellar medium (ISM). 
This in fact drops basic assumption (4). However, numerical studies 
of  [24,20,36,37] show that the break of the light curve 
is weaker and smoother than the one analytically predicted when the light travel 
effects related to the lateral size of the jet and  a realistic expression of the 
lateral expansion speed are taken into account. In the case of a jet expanding 
in a wind, the calculated light curve is even much weaker and smoother than the 
ananlytical one [22,17]. We recently calculated light 
curves for GRB afterglows when anisotropic jets ($dE/d\Omega\propto 
\theta^{-k}$) expand both in the interstellar medium and in the wind medium
[5]. We found that in each type of medium, one break appears 
in the late-time afterglow light curve for small $k$ but becomes weaker and 
smoother as $k$ increases. When $k\ge 2$, 
the break seems to disappear but the afterglow decays 
rapidly. Thus, we expect that the emission from 
expanding, highly anisotropic jets provides a plausible 
explanation for some rapidly fading afteglows whose 
light curves seem to have no break. 

We discussed the dense medium model in more details [12], by 
taking into account both the synchrotron self-absorption effect in the shocked 
medium and the energy injection effect of [8,10]. 
Recently, we combined the dense medium model with the 
pulsar energy injection effect to provide a good fit to the optical afterglow 
data of GRB 000301C [13]. Here we want to give a brief review of some of 
our studies on GRB afterglows.
 
\section{Shock Evolution}
 
It is well known that the evolution of a partially radiative shock depends
on both the efficiency with which the shock transfers its bulk kinetic
energy to electrons and magnetic fields and on the efficiency with which
the electrons radiate their energy. In 1998, we proposed, for the first time, 
a {\em unified} model for dynamical evolution of a partially radiative shock 
[6]. This model is not only valid during the whole evolution stage including 
the Sedov phase for an adiabatic shock, but also can describe well  
an adiabatic shock as well as a highly radiatve shock. This model was later 
re-investigated and referred to as a {\em generic} one in [19]. For simplicity, 
we here assume that a relativistic shock expanding in a dense medium is 
adiabatic. This assumption is correct particularly for a low electron energy 
density fraction in the shocked medium [6,7]. The Blandford-McKee 
self-similar solution [1] gives the Lorentz factor of an adiabatic relativistic shock,
\begin{equation}
\gamma=\frac{1}{4}\left[ \frac{17E_0(1+z)^3}{\pi nm_pc^5t_\oplus^3}
                  \right]^{1/8}
      =1E_{52}^{1/8}n_5^{-1/8}t_\oplus^{-3/8}[(1+z)/2]^{3/8},
\end{equation}
where $E_0=E_{52}\times 10^{52}{\rm ergs}$ is the total isotropic energy,
$n_{5}=n/10^5\,{\rm cm}^{-3}$, $t_\oplus$ is the observer's time
since the gamma-ray trigger in units of 1 day, $z$ is the the redshift
of the source generating this shock, and $m_p$ is the proton mass.
We assume $\gamma=1$ when $t_\oplus=t_b$. This implies
\begin{equation}
n_5=E_{52}t_b^{-3}[(1+z)/2]^3.
\end{equation}
For $t_\oplus > t_b$, the shock will be in a nonrelativistic phase. In the
following we will discuss the spectrum and light curve during the 
non-relativistic phase.  

As usual, only synchrotron radiation from the shock is considered.
To analyze the spectrum and light curve, one needs to know three
crucial frequencies: the synchrotron peak frequency ($\nu_m$), the cooling
frequency ($\nu_c$), and the self-absorption frequency ($\nu_a$).
We assume a power law distribution of the electrons accelerated
by the shock: $dn'_e/d\gamma_e\propto \gamma_e^{-p}$ for $\gamma_e
\ge\gamma_{em}$, where $\gamma_e$ is the electron Lorentz factor and
$\gamma_{em}=610\epsilon_e(\gamma-1)$ is the minimum Lorentz factor.
We further assume that $\epsilon_e$ and $\epsilon_B$ are the electron
and magnetic energy density fractions of the shocked medium respectively.
The $\nu_m$ is the characteristic synchrotron frequency of an electron
with Lorentz factor of $\gamma_{em}$, while the $\nu_c$ is the characteristic
synchrotron frequency of an electron which cools on the dynamical age of
the shock. According to Sari et al. [32], we have derived 
the synchrotron peak frequency, the cooling frequency and the synchotron 
self-absorption frequency, measured in the observer's frame [12].  
They are correct for the whole evolution stage. 
 
Now we give the spectrum and light curve of the afterglow during the 
non-relativistic phase. First, for the case without energy injection, the shock 
velocity decays as $\propto t_\oplus^{-3/5}$ and thus we have 
\begin{equation}
F_\nu=\left \{
       \begin{array}{lll}
         (\nu_a/\nu_m)^{-(p-1)/2}(\nu/\nu_a)^{5/2}F_{\nu_m}
         \propto \nu^{5/2}t_\oplus^{11/10} & {\rm if}\,\, \nu<\nu_a \\
         (\nu/\nu_m)^{-(p-1)/2}F_{\nu_m} \propto \nu^{-(p-1)/2}
         t_\oplus^{(21-15p)/10} & {\rm if}\,\, \nu_a< \nu <\nu_c \\
         (\nu_c/\nu_m)^{-(p-1)/2}(\nu/\nu_c)^{-p/2}F_{\nu_m}\propto
              \nu^{-p/2}t_\oplus^{(4-3p)/2} & {\rm if}\,\, \nu>\nu_c.
        \end{array}
       \right.
\end{equation}
We easily see that for high-frequency radiation the temporal decay index
$\alpha=(21-15p)/10$ for emission from slow-cooling electrons or
$\alpha=(4-3p)/2$ for emission from fast-cooling electrons.
If $p\approx 2.8$, then $\alpha\approx -2.1$ or $-2.2$. Comparing
this with the relativistic result, we conclude that the afterglow decay
steepens at the nonrelativistic stage.

Some models for GRB energy sources (for a brief review see [10])
predict that during the formation of
an ultrarelativistic fireball required by GRB, a strongly magnetized
millisecond pulsar will be born. If so, the pulsar will continuously input
its rotational energy into the forward shock of the postburst fireball
through magnetic dipole radiation because electromagnetic waves radiated
by the pulsar will be absorbed in the shocked medium [8,10]. 
Since an initially ultrarelativistic shock discussed in [12]
rapidly becomes nonrelativistic in a dense medium, we next investigate
the evolution of a nonrelativistic adiabatic shock with energy injection
from a pulsar. The total energy of the shock is the sum of the initial
energy and the energy which the shock has obtained from the pulsar:
\begin{equation}
E_0+\int_0^{t_\oplus}Ldt_\oplus =E_{\rm tot} \propto v^2r^3,
\end{equation}
where $L$ is the stellar spindown power $\propto (1+t_\oplus/T)^{-2}$
($T$ is the initial spindown time scale). The term on the right-hand side 
is consistent with the Sedov solution. Please note that $L$ can be thought of as
a constant for $t_\oplus<T$, while $L$ decays as $\propto t_\oplus^{-2}$
for $t_\oplus\gg T$. Because of this feature, we easily integrate the
second term on the left-hand side of equation (17). We now define a time
at which the shock has obtained energy $\sim E_0$ from the pulsar,
$t_c=E_0/L$, and assume $t_c\ll T$. The evolution of the
afterglow from such a shock can be divided into three stages.

Stage (i): $t_\oplus\ll t_c$, viz., the second term
on the left-hand side of equation (4) can be neglected. The evolution of
the afterglow is the same as in the above case without any energy injection.

Stage (ii): for $T>t_\oplus\gg t_c$, the term $E_0$ in equation (4)
can be neglected. At this stage, the shock's velocity $v\propto t_\oplus^{-2/5}$,  
we have derived the spectrum and light curve of the afterglow [12]
\begin{equation}
F_\nu=\left \{
       \begin{array}{lll}
         (\nu_a/\nu_m)^{-(p-1)/2}(\nu/\nu_a)^{5/2}F_{\nu_m}
         \propto \nu^{5/2}t_\oplus^{7/5} & {\rm if}\,\, \nu<\nu_a \\
         (\nu/\nu_m)^{-(p-1)/2}F_{\nu_m} \propto \nu^{-(p-1)/2}
         t_\oplus^{(12-5p)/5} & {\rm if}\,\, \nu_a< \nu <\nu_c \\
         (\nu_c/\nu_m)^{-(p-1)/2}(\nu/\nu_c)^{-p/2}F_{\nu_m}\propto
              \nu^{-p/2}t_\oplus^{2-p} & {\rm if}\,\, \nu>\nu_c.
        \end{array}
       \right.
\end{equation}
It can be seen that for high-frequency radiation the temporal decay index
$\alpha=(12-5p)/5\approx -0.4$ for emission from slow-cooling electrons
or $\alpha=2-p\approx -0.8$ for emission from fast-cooling electrons
if $p\approx 2.8$. This shows that the afterglow decay may significantly
flatten due to the effect of the pulsar.

Stage (iii): for $t_\oplus\gg T$, the power of the pulsar due to magnetic dipole
radiation rapidly decreases as $L\propto t_\oplus^{-2}$, and the evolution
of the shock is hardly affected by the stellar radiation. Thus,
the evolution of the afterglow at this stage will be the same as in
the above case without any energy injection.

In summary, as an adiabatic shock expands in a dense medium
from an ultrarelativistic phase to a nonrelativistic phase, the decay
of radiation from such a shock will steepen, subsequently may flatten
if a strongly magnetic millisecond pulsar continuously inputs
its rotational energy into the shock through magnetic dipole radiation,
and finally the decay will steepen again due to disappearance of the stellar
effect. In the next section, we will see how to explain some unusual afterglows
based on the above conclusion. 

\section{Some Unusual Afterglows}
 
\subsection{GRB 980519}

The optical afterglow $\sim 8.5$ hours after GRB 980519 decayed 
as $\propto t_\oplus^{-2.05\pm 0.04}$ in {\em BVRI} [18],  while 
the power-law decay index of the X-ray afterglow 
$\alpha_X=2.07\pm 0.11$ [25], in agreement
with the optical. The spectrum in optical band alone is well fitted by
a power low $\nu^{-1.20\pm 0.25}$, while the optical and X-ray spectra
together can also be fitted by a single power law of the form
$\nu^{-1.05\pm 0.10}$. In addition, the radio afterglow of this burst
was observed by the VLA at 8.3 GHz, and its temporal evolution
$\propto t_\oplus^{0.9\pm 0.3}$ between 1998 May 19.8UT and 
22.3UT [14].

We now analyze the observed afterglow data of GRB 980519 based on our
model. We assume that for this burst, the forward shock evolved from
an ultrarelativistic phase to a nonrelativistic phase in a dense
medium at $\sim 8$ hr after the burst. So, the detected afterglow,
in fact, was the radiation from a nonrelativistic shock. This implies
$\gamma\sim 1$ at $t_b\approx 1/3$ days. From equation (2), therefore,
we find
\begin{equation}
n_5\sim 27E_{52}[(1+z)/2]^3.
\end{equation}
If $p\approx 2.8$, and if the observed optical afterglow was emitted by
slow-cooling electrons and the X-ray afterglow from fast-cooling
electrons, then according to equation (3), the decay index $\alpha_R
=(21-15p)/10\approx -2.1$ and $\alpha_X=(4-3p)/2\approx -2.2$,
in excellent agreement with observations. Furthermore, the model spectral
index at the optical to X-ray band and the decay index at the
radio band, $\beta=-(p-1)/2 \approx -0.9$ and $\alpha=1.1$, are quite
consistent with the observed ones, $-1.05\pm 0.10$ and $0.9\pm 0.3$,
respectively.

We [12] took into account three observed data which correpond
to the radio, R-band and X-ray frequencies respectively, and inferred intrinsic 
parameters of the shock and the redshift of the burst,
$\epsilon_e\sim 0.16$, $\epsilon_B\sim 2.8\times 10^{-4}$, $E_{52}\sim 0.27$,
$n_5\sim 3.4$, and $z\sim 0.55$.
After considering these reasonable parameters, we numerically 
studied the trans-relativistic evolution of the shock [35] and found 
that our dense medium model can provide an excellent fit to all the observational 
data of the radio afterglow from GRB 980519 shown in [15]. 

\subsection{GRB 000301C}

The optical afterglow data of GRB 000301c were presented in [30]. 
In addition, the spectral index $\beta=1.1\pm 0.1$. 
We [13] combined the dense medium model
with the pulsar energy injection effect to explain the unusual optical afterglow. 
For stage (ii),  if $p=3.4$, then $\alpha=(12-5p)/5=-1.0$ and $\beta=-(p-1)/2=-1.2$ 
are consistent with the GRB 000301c R-band afterglow data in initial 7.5 days 
after the burst. These data indicate $\alpha_1\sim -1.1$, which implies 
$\alpha_{\rm obs}\sim \beta_{\rm obs}$ at early times. If the afterglow 
were radiated by fast-cooling electrons in the shocked medium, we would find 
$\alpha=2(1-\beta)$, which is clearly inconsistent with the observational 
result. Therefore, the GRB 000301c R-band afterglow arose from those 
slow-cooling electrons in the shocked medium. For stage (iii), in the case of 
$p=3.4$, the model's time index $\alpha=(21-15p)/10=-3.0$ is quite consistent 
with the observational data of the GRB 000301c R-band afterglow at late times, 
$\alpha_2=-3.01\pm 0.53$ [30]. We also carried out simulations of 
the evolution of a shock with energy injection from a pulsar and the resulting 
emission [33]. Our numerical results indeed show one sharp 
break in the late-time afterglow light curve and give a good fit to the R-band 
afterglow data of GRB 000301c.  

\section{Conclusions}
 
We discussed the evolution of an adiabatic shock 
expanding in a dense medium from an ultrarelativistic phase
to a nonrelativistic phase in more details in this paper. In particular,
we discussed the effects of synchrotron self absorption and energy
injection on the afterglow from this shock. In a dense medium, the shock
becomes nonrelativistic rapidly after a short relativistic phase.
This transition time varies from several hours to a few days when
the medium density is from $10^5$ to a few $\times 10^6\,{\rm cm}^{-3}$,
and the shock energy from $10^{51}$ to $10^{54}$ ergs.
The afterglow from the shock at the nonrelativistic stage 
decays more rapidly than at the relativistic stage, while the decay
index varies from $-1.35$ to $-2.1$ if the spectral index of
the accelerated electron distribution, $p=2.8$, and
the radiation comes from those slow-cooling electrons. Since some models
mentioned above predict that a strongly magnetic millisecond pulsar
may be born during the formation of GRB, we also discuss the effect of such
a pulsar on the evolution of a nonrelativistic shock through magnetic dipole
radiation, in contrast to the case discussed in [8,10].
We found that after the energy which the shock obtains from the pulsar
is much more than the initial energy of the shock, the afterglow decay
will flatten significantly and the decay index will become $-0.4$.
When the pulsar energy input effect disappears, the index will still be $-2.1$.
These features are in excellent agreement with the afterglow  of
GRB 980519. Furthermore, our model fits very well
all the observational data of GRB 980519 including all the radio data.
Our model also provides a good fit to the R-band afterglow data of GRB 000301c.  

\subsection*{Acknowledgments}
We would like to thank T. Lu, D. M. Wei and Y. F. Huang for helpful discussions.
This work was supported by the National Natural Science 
Foundation of China (grant 19825109) and by 
the National 973 Project.


\begin{thebibliography}{99}

\bibitem {} Blandford, R. D., \& McKee, C. F., Phys. Fluids {\bf 19}, 1130 (1976).
\bibitem {} Castro-Tirado, A. J. et al., Science {\bf 283}, 2069 (1999).
\bibitem {} Chevalier, R. A., \& Li, Z. Y., ApJ {\bf 520}, L29 (1999).
\bibitem {} Chevalier, R. A., \& Li, Z. Y., ApJ {\bf 536}, 195 (2000).
\bibitem {} Dai, Z. G.,  \& Gou, L. J., ApJ, submitted (2000).
\bibitem {} Dai, Z. G., Huang, Y. F., \& Lu, T., preprint: astro-ph/9806334.
\bibitem {} Dai, Z. G., Huang, Y. F., \& Lu, T., ApJ {\bf 520}, 634 (1999). 
\bibitem {} Dai, Z. G., \& Lu, T., Phys. Rev. Lett. {\bf 81}, 4301 (1998).
\bibitem {} Dai, Z. G., \& Lu, T., MNRAS {\bf 298}, 87 (1998).
\bibitem {} Dai, Z. G., \& Lu, T., A\&A {\bf 333}, L87 (1998).
\bibitem {} Dai, Z. G., \& Lu, T., ApJ {\bf 519}, L155 (1999). 
\bibitem {} Dai, Z. G., \& Lu, T., ApJ {\bf 537}, 803 (2000).
\bibitem {} Dai, Z. G., \& Lu, T., A\&A, submitted (astroph/0005417).
\bibitem {} Frail, D. A., Taylor, G. B., \& Kulkarni, S. R., GCNC {\bf 89} (1998).
\bibitem {} Frail, D. A., et al., preprint: astro-ph/9910060.
\bibitem {} Fruchter, A. S. et al., preprint: astro-ph/9902236.
\bibitem {} Gou, L. J. Dai, Z. G., Huang, Y. F., \& Lu, T., A\&A, submitted (2000). 
\bibitem {} Halpern, J. P., Kemp, J., Piran, T., \& Bershady, M. A.,
          ApJ {\bf 517}, L105 (1999).
\bibitem {} Huang, Y. F., Dai, Z. G., \& Lu, T., MNRAS {\bf 309}, 513 (1999).
\bibitem {} Huang, Y. F., Gou, L. J., Dai, Z. G., \& Lu, T., ApJ, in press (2000).
\bibitem {} Kulkarni, S. R. et al., Nature {\bf 398}, 389 (1999).
\bibitem {} Kumar, P., \& Panaitescu, A., preprint: astro-ph/0003264. 
\bibitem {} M\'esz\'aros, P., Rees, M. J., \& Wijers, R. A. M. J., ApJ, 499, 301
\bibitem {} Moderski, R., Sikora, M., \& Bulik, T., ApJ {\bf 529}, 151 (2000).
\bibitem {} Owens, A. et al., A\&A {\bf 339}, L37 (1998).
\bibitem {} Panaitescu, A., M\'esz\'aros, P., \& Rees, M. J., ApJ {\bf 503}, 315 (1998).
\bibitem {} Piran, T., Phys. Rep. {\bf 314}, 575 (1999).
\bibitem {} Rees, M. J., \& M\'esz\'aros, P., ApJ {\bf 496}, L1 (1998).
\bibitem {} Rhoads, J., ApJ {\bf 525}, 737 (1999).
\bibitem {} Sagar, R. et al., preprint: astro-ph/0004223.
\bibitem {} Sari, R., Piran, T., \& Halpern, J. P., ApJ {\bf 519}, L17 (1999).
\bibitem {} Sari, R., Piran, T., \& Narayan, R., ApJ {\bf 497}, L17 (1999).
\bibitem {} Wang, W., \& Dai, Z. G., ApJ, submitted (2000).
\bibitem {} Wang, X. Y., Dai, Z. G., \& Lu, T., MNRAS, in press (2000).
\bibitem {} Wang, X. Y., Dai, Z. G., \& Lu, T., MNRAS, in press (2000).
\bibitem {} Wei, D. M., in these proceedings.
\bibitem {} Wei, D. M., \& Lu, T., ApJ, in press (2000).
\bibitem {} Wijers, R. A. M. J., in these proceedings.
\end{thebibliography}
\end{document}